\documentclass[a4paper,conference]{IEEEtran}

\ifCLASSINFOpdf

\else

\fi

\usepackage{graphicx}
\usepackage{amsmath}
\usepackage{amssymb}
\usepackage[compress]{cite}
\usepackage{color}
\usepackage{subfigure}

\newtheorem{prop}{Proposition}
\newtheorem{rem}{Remark}

\newcommand{\figsizee}{0.38}

\begin{document}

\title{Hybrid RF/VLC Systems under QoS Constraints}

\author{\IEEEauthorblockN{Marwan Hammouda\IEEEauthorrefmark{1}, Sami Ak{\i}n\IEEEauthorrefmark{1},
Anna Maria Vegni\IEEEauthorrefmark{2}, Harald Haas\IEEEauthorrefmark{3}, and J\"{u}rgen Peissig\IEEEauthorrefmark{1}}\\
\IEEEauthorblockA{
\IEEEauthorrefmark{1}Institute of Communications Technology, Leibniz Universit\"{a}t Hannover, Hannover, Germany \\ Email: \{marwan.hammouda, sami.akin, and peissig\}@ikt.uni-hannover.de\\
\IEEEauthorrefmark{2}Department of Engineering, COMLAB Telecommunications Lab, Roma Tre University, Rome, Italy\\
Email: annamaria.vegni@uniroma3.it\\
\IEEEauthorrefmark{3}Li-Fi R\&D Centre, Institute for Digital Communications, University of Edinburgh, Edinburgh, EH9 3JL, UK.\\
Email: h.haas@ed.ac.uk}}

\maketitle

\begin{abstract}
The coexistence of radio frequency (RF) and visible light communications (VLC) in typical indoor environments can be leveraged to support 
vast user quality-of-service (QoS) needs. In this paper, we target a hybrid RF/VLC network in which data transmission is provided via either 
an RF access point or a VLC luminary based on a selection process. We employ the concept of effective capacity, which defines the maximum 
constant arrival data rate at the transmitter buffer when the QoS needs are imposed as limits on the buffer overflow and delay violation probabilities, as the main selection criteria. 
We initially formulate the effective capacity of both channels with respect to channel gains and user distribution. Under the assumption of 
uniform user distribution within the VLC cell, we then provide a closed-form approximation for the effective capacity of the VLC channel. 
We further investigate the effects of illumination needs and  line-of-sight  blockage on the VLC performance. In addition, we 
explore the non-asymptotic bounds regarding the buffering delay by capitalizing on the effective capacity. Through simulation results, 
we show the impacts of different physical aspects and data-link QoS needs on the effective capacity and  delay bounds.   
\end{abstract}


\section{Introduction}
Hybrid networks that integrate radio frequency (RF) and visible light communication (VLC) technologies have been proposed intensively in the recent years in order to achieve the end-user demands of both capacity and coverage, which are difficult to meet when either technology is operating solely. In addition, such networks can be practically feasible with no extra infrastructure costs since both RF and VLC systems already coexist and operate in the same area in many indoor scenarios, like offices.  In this line of research, the authors in~\cite{basnayaka2015hybrid,rahaim2011hybrid, kashef2016energy,basnayaka2017design} explored the performance of hybrid RF/VLC systems, where the results showed substantial improvements over pure RF and pure VLC networks in terms of throughput and energy efficiency. To support user mobility in hybrid RF/VLC systems, the authors in~\cite{wang2015efficient, wang2015dynamic} proposed different switching algorithms between both technologies to ensure a seamless handover, thus maintaining connectivity. Furthermore, various load balancing schemes in hybrid RF/VLC systems were addressed in \cite{wang2017optimization,li2015cooperative}.

All of the these studies focused mainly on the physical layer aspects of the hybrid RF/VLC systems. However, the increasing demand on delay-sensitive applications, such as games and video streaming, requires involving other quality-of-service (QoS) metrics at the data-link layer. Subsequently, cross-layer analyses regarding the physical and data-link layers were addressed by many researchers when the QoS constraints become necessary. In this direction, effective capacity was proposed by Wu and Negi \cite{wu2003effective} as a cross-layer performance metric, which identifies the maximum {\it constant} arrival rate that a given service (channel) process can sustain under statistical QoS requirements imposed as delay and buffer violation probabilities. Since then, the effective capacity has been gaining an increasing attention in the RF literature, and it has been investigated in different RF transmission scenarios~\cite{tang2007quality, hammouda2014effective,hammouda2016effective,akin2010effective}. Nevertheless, channel randomness due to user distribution is not considered in these studies, and only small-scale channel fading is assumed. On the other hand, to the best of our knowledge, there are only few studies that recently investigated the effective capacity in VLC systems\cite{jin2016resource,jin2015resource}.

In this paper, we provide a cross-layer study of a hybrid RF/VLC system that operates under statistical QoS constraints, which are inflicted as limits on the buffer overflow and delay violation probabilities. We further consider a random user location. Particularly, we consider a hybrid RF/VLC system in which data transmission is performed over either the RF link or the VLC link following a selection process and we utilize the effective capacity concept as the main link selection criteria. Different than the studies in \cite{jin2016resource,jin2015resource}, \textit{i}) we provide a closed-form approximation for the effective capacity of the VLC link assuming that the user is uniformly located within the VLC coverage area, \textit{ii}) we further exploit the effective capacity principle to explore the non-asymptotic bounds regarding the buffering delay when either the RF or VLC channel is selected for data transmission, and \textit{iii}) we study the impacts of illumination requirements and  different physical aspects, including channel statistics and transmission patterns on the system performance. 


\section{System Model}
\label{sec:System_Model}
\begin{figure}
\centering
\includegraphics[width=\figsizee\textwidth]{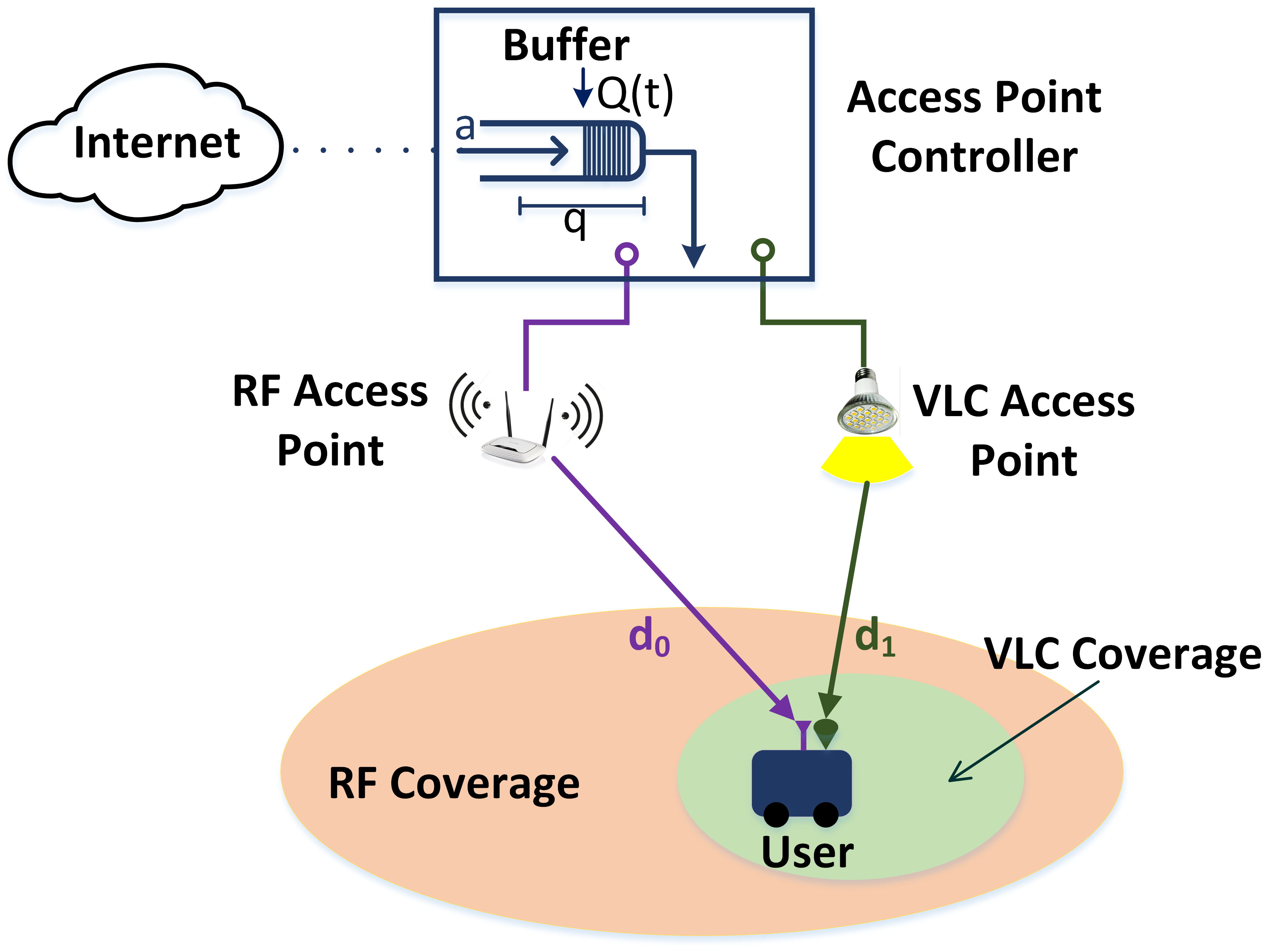}
\caption{Model of the hybrid RF/VLC indoor system.}
\label{Channel_Model}
\end{figure}

As depicted in \figurename~\ref{Channel_Model}, we consider a hybrid RF/VLC system in which one RF and one VLC access points\footnote{Notice that the analytical framework provided in this paper can be directly applied to general scenarios with multiple VLC APs when interference mitigation techniques are employed, see e.g., \cite{hammouda2017design}. Nevertheless, an extension to interference scenarios can be easily formulated.} (APs) are connected to an access point controller. We assume that the user is uniformly located within the coverage area of the VLC AP, which is also covered by the RF AP, and that the user~\footnote{In this paper, we use the terms {\it user} and {\it receiver} interchangeably.} is  equipped with both RF and VLC front-ends. Nevertheless, we assume that the user does not have a multi-homing capability to perform link aggregation, and hence it can receive data over only one link, either RF or VLC, at any time instant. Herein, we assume that the AP controller is responsible for selecting the most appropriate link for data transmission based on a certain selection criteria. In particular, we consider a point-to-point downlink scenario and assume that a transmission link is established between the access point controller and the user through either the RF AP or the VLC AP at any time instant. The data initially arrive at the access point controller from a source (or sources) and are stored in the buffer before being divided into packets and conveyed into the selected channel in frame of $T$ seconds. Notice that the two APs are located at different positions. In particular, the distance between the user and the RF AP is denoted as $d_0$, whereas the distance between the user and the RF AP is denoted as $d_1$, as shown in \figurename~\ref{Channel_Model}. In the following two subsections, we introduce the RF and the VLC channel models, respectively.
\subsection{RF Channel Model}
When the flat-fading RF channel is selected for data transmission, the input-output relation 
at time instant $t$ is given by
\begin{equation}
\label{input_output_RF}
y_r(t) = x_{r}(t) h_r(t) + n_{r}(t), 
\end{equation}
where $x_{r}(t)$ and $y_r(t)$ are, respectively, the complex input and output of the RF channel input. Herein, the channel input is assumed to be subject to the following average power constraint: $\mathbb{E}\{|x_{r}(t)|^2\} \leq P_{r},$
where $P_{r}$ is the maximum allowed average power in the RF link. Moreover, $n_{r}(t)$  is the additive thermal noise at the RF front-end of the receiver, which is a zero-mean, circularly symmetric, complex Gaussian random variable with a 
variance $\sigma_{r}^2$, \textit{i.e.}, $\mathbb{E}\{|n_{r}(t)|^2\} = \sigma_{r}^2$. The noise samples $\{n_{r}(t)\}$ are assumed to be independent and identically distributed. Also, $h_{r}(t)$ is the fading coefficient of the RF channel, which has an arbitrary distribution with a finite variance, \textit{i.e.}, $\mathbb{E}\{|h_{r}(t)|^2\} \leq \infty$. We consider a block-fading channel and assume that $h_{r}(t)$ stays fixed during one transmission frame, \textit{i.e.}, $T$ seconds, and changes independently from one frame to another. We further underline that the available bandwidth for the RF channel is $B_{r}$ [Hz], and hence the symbol rate is equal to $T B_{r}$ complex symbols in each transmission frame. Consequently, the achievable rate (capacity lower bounds) for the RF channel in the $l^{\text{th}}$ time frame is given in {\it bits per frame} by~\cite{shamai1995capacity}:
\begin{equation}
\label{Rate_RF}
R_{r,l} = T B_r \log_2 \bigg(1+\frac{P_{r} |h_{r,l}|^2}{\sigma_{r}^2}\bigg),
\end{equation} 
where $h_{r,l}$ is the channel fading gain in the $l^{\text{th}}$ frame.
\subsection{VLC Channel Model}
We assume that the simple and widely used scheme of intensity modulation and direct detection (IM/DD) is employed. In IM/DD, the transmitting light emitting diode (LED) varies the emitted light intensity  with respect to the transmitted signal. On the other hand, the VLC front-end of the receiver is equipped with a photodetector (PD) which generates an electrical current (or voltage)  proportional to the collected light intensity. VLC channels are typically composed of both  Line-of-Sight (LoS) and diffuse (multi-path) components. However,  in~\cite{komine2004fundamental} it was observed that, in typical indoor scenarios the majority of the collected energy at the photodetector (more than $95\%$) comes from the LoS component. Therefore, in this paper we mainly assume that the VLC link has a dominant LoS component. In such a case, the VLC channel can be considered 
flat~\cite{pohl2000channel}, and the received signal at the PD at time instant $t$, \textit{i.e.}, $y_v(t)$,
is given by~\cite{yin2015performance}
\begin{equation}
\label{input_output_VLC}
y_v(t) = x_{v}(t) h_v + n_{v}(t), 
\end{equation}
where $x_{v} \in \mathbb{R}^{+}$ is the emitted intensity by the LED, whose average value is upper bounded as $\mathbb{E}\{x_v\} = P_v$ due to safety concerns \footnote{Notice that a peak intensity constraint can also be imposed for practical and safety concerns. However, we ignore such a limit in this paper for the sake of simplicity.}. Moreover, $h_{v} \in \mathbb{R}^{+}$ is the optical channel gain, which is time-invariant and depends only on the user position, as will be detailed in the sequel. In (\ref{input_output_VLC}), $n_{v}(t)$ is the additive thermal noise at the PD front-end, which is a real-valued Gaussian random variable with zero mean and variance $\sigma_{v}^2$. The noise samples $\{n_{v}\}$ are further assumed to be independent and identically distributed. Notice that, unlike the channel input $x_v(t)$, the channel output $y_v(t)$ may be negative due to the noise samples~\cite{lapidoth2009capacity}. 

\begin{figure}
\centering
\includegraphics[width=0.3\textwidth]{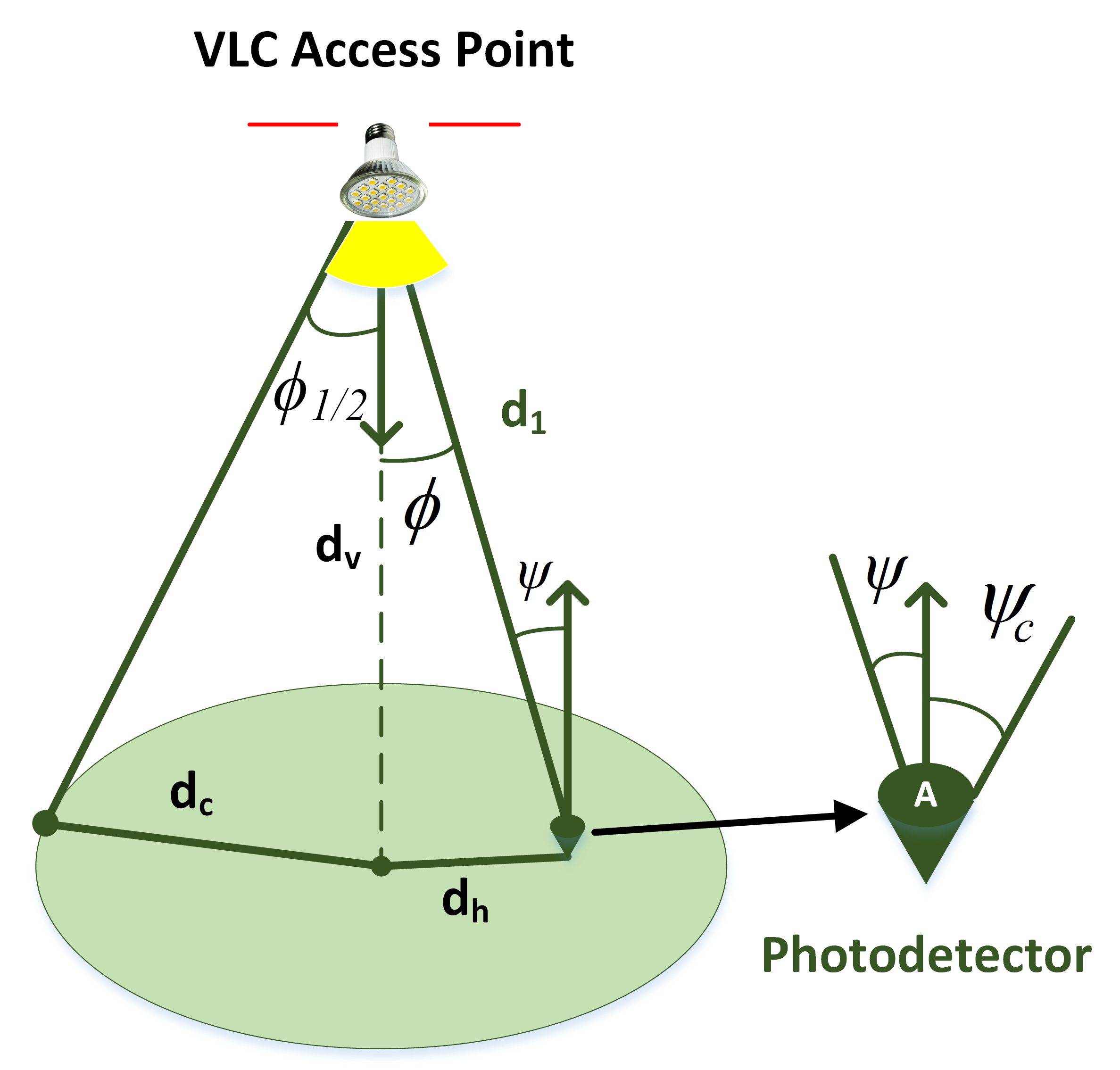}
\caption{VLC channel via LoS link.}
\label{VLC_LOS}
\end{figure}

More in detail, in \figurename~\ref{VLC_LOS} we depict a VLC link via LoS connection. We assume that the LED-based AP follows the Lambertian radiation pattern, the AP is directed downwards, and the user PD is directed upwards. Thus, the channel gain at a given distance $d_1$ between the AP and the user is expressed as~\cite{barry1993simulation}
\begin{equation}
\label{VLC_gain_1}
h_{v} = \frac{(r+1)A D(\psi) n^2 d_v^{r+1}}{2 \pi \sin^2(\psi_C) d_1^{r+3}}\text{rect}(\psi/\psi_C),
\end{equation}
where $A$, $\psi$, and $\psi_C$ are, respectively, the PD physical area, the angle of incidence with respect to the normal axis to the receiver plane, and the field of view (FOV) angle of the photodetector. In addition, $D(\psi)$ is the gain of the optical filter, and $n$ is the refractive index. In \eqref{VLC_gain_1}, $r = -1/\log_2(\cos(\phi_{1/2}))$ is the Lambertian index, where $\phi_{1/2}$ is the LED half intensity viewing angle. Further, $\text{rect}(z)$ is an indicator function such that $\text{rect}(z) = 1$ if $z \leq 1$ and $\text{rect}(z) = 0$ otherwise.

Following the optical-to-electrical conversion, the signal-to-noise ratio at the VLC receiver can be defined as follows~\cite{wang2015dynamic}:
\begin{equation}
\label{SNR_vlc}
\zeta_{v} = \frac{ (\alpha P_{v} h_{v})^2}{\varsigma^2 \sigma_{v}^2},
\end{equation}
where $\alpha$ is the optical-to-electrical conversion efficiency of the PD and $\varsigma$ is the ratio between the average optical power and the average electrical power of the transmitted signal. Setting $\varsigma = 3$ can guarantee a negligible clipping noise, and hence the LED can be considered to be working in its linear region. Herein, we consider the achievable rate in \textit{bits per frame} for the VLC link of the form \cite{chaaban2017fundamental}
\begin{equation}
\label{Rate_VLC}
R_v = \frac{T B_v}{2} \log_2(1+c^2 \zeta_v),
\end{equation}
for some constant $c$, where $B_v$ is the available bandwidth in the VLC channel. For instance, $c = \sqrt{e/2 \pi}$ when the transmitted light intensity is exponentially distributed~\cite{lapidoth2009capacity}.

It is important to remark that, while we consider a random (and uniform) user distribution within the VLC coverage area, we assume that the perfect knowledge of the selected channel, i.e., either RF or VLC, is available at the transmitter side in each frame. Therefore, we set the instantaneous transmission rate in the selected channel to the achievable rate in that channel, and hence a reliable transmission can be guaranteed. Notice that the selection process regarded in this paper is not be performed in a frame-by-frame basis, and it is not based on the user's exact location. Instead, we provide a system-level selection process, which mainly depends on the system structure and user distribution. This process also agrees with the considered selection criteria, i.e., effective capacity, which is a steady-state performance metric, as it will become clear in the next section.
\section{Effective Capacity}\label{sec:link_sel}
Recall that the AP controller holds the data in its buffer before being transmitted over the selected link, \textit{i.e.}, either the RF link or the VLC link. Thus, applying certain constraints on the buffer length is required in order to control delay and overflow probabilities. In detail, let $Q$ be the steady-state buffer length and $q$ be a given threshold. Herein, we are interested in the QoS constraints regarding the buffer violation probability, \textit{i.e.}, $\text{Pr} \{ Q \geq q\}$, in the form of

\begin{equation}
\label{buffer_limit}
\theta = - \lim_{q \to \infty} \frac{\log \text{Pr} \{ Q \geq q\}}{q},
\end{equation}
for a given $\theta>0$, which represents the decay rate of the tail distribution of the queue length. Following~\eqref{buffer_limit}, we can approximate the buffer violation probability for a large threshold, \textit{i.e.}, $q_{\text{max}}$, as $\text{Pr}(Q \geq q_{max}) \approx e^{-\theta q_{max}}.$ This expression implies that, for a large threshold, the buffer violation probability decays exponentially with a rate controlled by $\theta$, which is also denoted as the QoS exponent. In particular, larger $\theta$ implies stricter QoS constraints, whereas smaller $\theta$ corresponds to looser constraints.

In this paper, we focus on the data arrival process at the buffer and  we employ the effective capacity, which defines the maximum {\it constant} arrival rate that a given service (channel) process can sustain while satisfying the limit in \eqref{buffer_limit}~\cite{wu2003effective}, as the main performance measure. For a given service process with a discrete-time, stationary
and ergodic stochastic service process $r(l)$ for $l = 1,2,\dots$, and for a given QoS exponent $\theta$, the effective capacity is formally defined as follows:
\begin{equation}
\label{eq:EC_gen}
C(\theta) = - \lim_{t \to \infty} \frac{1}{\theta t} \log_e \mathbb{E}\{e^{-\theta \sum_{l=1}^t r(l)}\}.
\end{equation}
Consequently, under the block-fading assumption of the RF and VLC channels, the effective capacity of the RF and VLC channels in \textit{bits per frame} can be, respectively, expressed as 

\begin{equation}
\label{EC_RF}
C_{r}(\theta) = -\frac{1}{\theta} \log_e \mathbb{E}_{h_r,d_h}\{e^{-\theta  R_{r,l}}\}, 
\end{equation} 
and
\begin{equation}
\label{EC_VLC}
C_{v}(\theta) = -\frac{1}{\theta} \log_e \mathbb{E}_{d_h}\{e^{-\theta R_{v}}\},
\end{equation} 
where $R_{r,l}$ and $R_v$ are given in \eqref{Rate_RF} and \eqref{Rate_VLC}, respectively. Notice that the expectation operation in \eqref{EC_RF} is generally performed with respect to both channel fading and user distribution, which is expressed in terms of the horizontal distance $d_h$.
 On the other hand, the expectation in \eqref{EC_VLC} is performed with respect to the user distribution only since the optical channel gain is time-invariant for a given user location\footnote{Small-scale variations in VLC channels (i.e., fading) is mitigated since
the area of a photodetector is much larger than the light wavelength [35, Sec. 2.5]. Thus, VLC channels are known as time-invariant.}. In the following preposition, we provide a closed-form approximation for the moment generation function $\mathbb{E}_{d_h}\{e^{-\theta R_{v}}\}$ in \eqref{EC_VLC} when the user is uniformly located within the VLC cell.

\begin{prop} \label{prop_1}
Let a user be uniformly located within the coverage area of a VLC AP with a radius $d_c$. Then, the moment generation function $\mathbb{E}_{d_h}\{e^{-\theta R_{v}}\}$ can be approximated as follows:
\begin{equation}
\label{MGF_VLC}
\begin{aligned}
&\mathbb{E}_{d_h}\{e^{-\theta R_{v}}\} \approx \\
&\frac{(\rho \omega)^{-\kappa}}{d_c^2 \kappa (r+3) + d_c^2} \bigg [ (d_c^2 + d_v^2)^{\kappa (r+3) + 1} - d_v^{2 \kappa (r+3) + 2}\bigg],
\end{aligned}
\end{equation}
where $\rho = \frac{c^2 P_{v}^2 \alpha^2}{\varsigma^2 \sigma_{v}^2}$, $\kappa = \frac{\theta T B_{v}}{2 \log_e(2)}$, and $\omega = \frac{A D(\psi) n^2}{2 \pi \sin^2(\psi_C)}$.
\end{prop}

\vspace{2mm}
\emph{Proof:} Notice that the distance $d_1$ can be expressed in terms of the horizontal distance between the user and the cell center, $d_h$, and the vertical distance between the transmitting and receiving planes, $d_v$, as $d_1 = \sqrt{d_h^2 + d_v^2}$. Then, the VLC channel gain, \textit{i.e.,} $h_v$ in (\ref{VLC_gain_1}), can be expressed as
\begin{equation}
h_v = \frac{(r+1)A D(\psi) n^2 d_v^{r+1}}{2 \pi \sin^2(\psi_C) (d_h^2 + d_v^2)^{\frac{r+3}{2}}}\text{rect}(\psi/\psi_C).
\end{equation}
 Under the assumption of user uniform distribution within the VLC cell, we consider $f_{d_h}(h) = 2h/d_c^2$ as the  probability density function (pdf) of the horizontal distance $d_h$. Consequently, the pdf of the square of the VLC channel gain, \textit{i.e.}, $h_v^2$, can be expressed as follows~\cite[Eq. (11)]{yin2015performance}:
\begin{equation*}
f_{h_v^2}(h) = \frac{1}{d_c^2}\frac{1}{r+3} [\omega(r+1)d_v]^{\frac{2}{r+3}} h^{\frac{-r-4}{r+3}}.
\end{equation*}
Then, we express the moment generation function $\mathbb{E}_{d_h}\{e^{-\theta R_{v}}\}$ in the integration form as 
\allowdisplaybreaks
\begin{align}
\mathbb{E}_{d_h}\{e^{-\theta R_{v}}\} & = \int  (1 + c^2 \zeta_v )^{-\frac{\theta T B_{v}}{2 \log_e(2)}} f_{h_v^2}(h) \mathrm{d}h \notag \\
& \approx\int  (c^2 \zeta_v )^{-\frac{\theta T B_{v}}{2 \log_e(2)}} f_{h_v^2}(h) \mathrm{d}h \label{proof_1}\\
& = \int \bigg (\frac{c^2 P_{v}^2 \alpha^2}{\varsigma^2 \sigma_{v}^2} h \bigg)^{-\frac{\theta T B_{v}}{2 \log_e(2)}} f_{h_v^2}(h) \mathrm{d}h \notag \\
& = \int_{\xi_{\min}}^{\xi_{\max}} \frac{(\rho h)^{-\kappa}}{h^{\frac{r+4}{r+3}}} \mathrm{d}h \label{proof_2},
\end{align}
where $\xi_{\min} = \frac{( \omega (r+1) d_v^{r+1})^2}{(d_v^2 + d_h^2)^{r+3}}$ and $\xi_{\max} = \frac{( \omega (r+1) d_v^{r+1})^2}{d_v^{2(r+3)}}$. The approximation in \eqref{proof_1} is based on the assumption that $\log(1+c^2 \zeta_v) \approx \log(c^2 \zeta_v)$, which is a reasonable assumption since the VLC channels generally have a high signal-to-noise ratio, \textit{i.e.}, $\zeta_v \gg 1$. Then, we solve the integration in \eqref{proof_2} to obtain the result in \eqref{MGF_VLC}. $\hfill{\square}$

On the other hand, the moment generation function of the RF link $\mathbb{E}_{h_r,d_h}\{e^{-\theta  R_{r,l}}\}$ in (\ref{EC_RF}) is hard to obtain in a closed-form due to the double-integration operation. Therefore, we resort to the Monte Carlo method in evaluating $\mathbb{E}_{h_r,d_h}\{e^{-\theta  R_{r,l}}\}$ in this paper. 

\begin{rem}
While we assume a dominant LoS component in the VLC link, the availability of such a link might be disturbed in some indoor environments, even when the user is located within the VLC cell. To address this issue, we model the LoS availability as a random event following the Bernoulli distribution with a success probability $\mu$, \textit{i.e.}, during each transmission frame the LoS link is available with probability $\mu$ and it is blocked (disturbed) with probability $1-\mu$. Noting that $R_v$ in \eqref{Rate_VLC} defines the achievable rate when the LoS is available, we assume that the achievable rate is $\Omega R_v$ otherwise, where $0 \leq \Omega < 1$. Assuming that the LoS availability is frame-wise independent, the effective capacity of the VLC channel in \eqref{EC_VLC} can be re-written as follows~\footnote{We refer to~\cite[Theorem 1]{akin2010effective} for detailed derivations concerning similar transmission settings and assumptions.}:
\begin{equation}
C_v(\theta) = -\frac{\log_e [\mu \mathbb{E}_{d_h}\{e^{-\theta R_{v}}\} + (1-\mu) \mathbb{E}_{d_h}\{e^{-\theta \Omega R_v}\}]}{\theta} .
\end{equation}
\end{rem}

\begin{rem}
Since the LED-based AP is originally employed for lighting purposes, the need for a sufficient amount of light over the receiving plane should be also considered. Notice that different indoor places may require different brightness levels based on the running activities, as defined by the  European lighting standard~\cite{Standard}. In this regard, illuminance, denoted as $E$, is the most commonly used measure that characterizes the brightness level at a given point. In this paper, we target a brightness span of $[E_{\text{min}},E_{\text{max}}]$ lx within the VLC cell coverage, such that the brightness level at the cell edge fulfills the minimum level of $E_{\text{min}}$, while it does not exceed $E_{\text{max}}$ at the cell center for the eye safety concerns. For such a case, we showed in~\cite{hammouda2017design} that the cell radius is limited as
\begin{equation}
\label{illumination_3}
d_{c}^2 \leq d_v^2 \bigg[\bigg(\frac{E_{\text{max}}}{E_{\text{min}}}\bigg)^{\frac{2}{r + 3}} - 1\bigg],
\end{equation}
or equivalently the LED viewing angle is limited as $\tan(\phi_{1/2}) \leq [(E_{\text{max}}/E_{\text{min}})^{\frac{2}{r + 3}} - 1]^{\frac{1}{2}}$. In this paper, we assume that the user is always located within the area that satisfies the lighting constraint in (\ref{illumination_3}). This assumption can be practically feasible if the VLC AP is equipped with a zooming capability that enables it to, physically, adjust its viewing angle based on the required illumination levels within the cell. Alternatively, the AP can allocate all transmission resources (\textit{i.e.}, power, time, and bandwidth) in the area that satisfies the illumination needs. This latter "cognitive" approach was proposed for more general settings in~\cite{hammouda2017design}.
\end{rem}

\paragraph*{Non-asymptotic Delay Bounds} \label{par:non_asym}
As observed from \eqref{eq:EC_gen}, the effective capacity is an asymptotic performance measure in time, \textit{i.e.}, the number of time frames is assumed to be infinitely large. However, non-asymptotic characterizations are of a high interest from practical perspectives. Herein, we target the statistical bound regarding the queuing delay at the transmitter, such that the stationary delay, $\mathcal{D}$, exceeds a given threshold, $d$, at most with a small probability, $\varepsilon$, \textit{i.e.}, $\text{Pr}\{\mathcal{D} > d\} \leq \varepsilon$. Let us assume a fixed arrival rate at the transmitter buffer of $a>0$ \textit{bits per frame}, and consider a first-come first-serve order. Thus, the delay bound corresponding to the RF channel can be expressed as follows~\cite[Eq. (24)]{fidler2015capacity}:
\begin{equation}
d = d_r = \frac{-\log(\theta(C_r(\theta)-a)\varepsilon)}{\theta a},
\end{equation}
where $\theta$ is a free parameter that satisfies
\begin{equation}
0 < \theta \leq \frac{1}{\varepsilon(C_r(\theta)-a)}.
\end{equation}
Notice that we have $C_r(\theta)>a$ for stability. Likewise, the delay bound corresponding to the VLC link is given by
\allowdisplaybreaks
\begin{equation}
\begin{aligned}
& d = d_v = \frac{-\log(\theta(C_v(\theta)-a)\varepsilon)}{\theta a},\\
& \text{for} \qquad 0 < \theta \leq \frac{1}{\varepsilon(C_v(\theta)-a)}.
\end{aligned}
\end{equation}
\section{Simulation Results}
\label{sec:sim}
In this section, we present the simulation results for the hybrid RF/VLC system model. Unless specified otherwise, we set the transmission frame period to 0.1 milliseconds, i.e., $T = 10^{-4}$. The thermal noise power of the RF front-end can be calculated as $\sigma_r^2 = N_r B_r$, and the thermal noise power of the PD as $\sigma_v^2 = N_v B_v$. Here, $N_r$ is the power spectral density of the RF front-end and $N_v$ is the power spectral density of the PD noise. Recalling that we consider a system-level selection process, in this section we mainly investigate the impacts of the geometry-related parameters, i.e., the VLC cell size and the locations of the two APs.

Regarding the RF channel, we consider a Rician fading distribution with a Rician factor $K$\footnote{Notice that we can also reflect the channel characteristics in millimeter wave range communications by properly setting the value of $K$\cite{basnayaka2017design}.}, such that the channel realizations $\{h_r(t)\}$ are independent and identically distributed complex Gaussian random variables with mean value and variance, respectively, as follows:
\begin{equation}
 \mu_h = \sqrt{\frac{e^{-L(d_0)/10} K}{K+1}}, \quad \text{and} \quad  \sigma_h^2 = \frac{e^{-L(d_0)/10}}{K+1}. 
\end{equation} 
Above, $L(d_0)$ is the large-scale path loss in decibels as a function of the distance between the RF AP and the receiver, $d_0$, which is given by~\cite{akl2006indoor} as 
\begin{equation}
L(d_0) = L ({d_{\text{ref}}}) + 10 \varrho \log\bigg(\frac{d_{0}}{d_{\text{ref}}}\bigg) + X_{\sigma},
\end{equation}
where $L ({d_{\text{ref}}}) = 40$~dB is the path loss at a reference distance $d_{\text{ref}} = 1$ m and an operating frequency of $2.4$~GHz. In addition, $\varrho$ is the path loss exponent and $X_{\sigma}$ represents the shadowing effect, which is assumed to a zero-mean Gaussian random variable with a standard deviation $\sigma$ expressed in decibels\footnote{Many experimental campaigns have been conducted, and empirical values of $K$, $\varrho$, and $\sigma$ for different indoor scenarios have been reported in several literature studies, see \textit{e.g.},~\cite{akl2006indoor}.}. Table~\ref{tab_1} summarizes the simulation parameters used in this paper\footnote{VLC system parameters are similar to those considered in \cite{basnayaka2017design}}. Let $(x_v, y_v, z_v) = (0,0,0)$ be the Cartesian coordinates of the VLC AP, $(x_r,y_r,z_r) = (0,y_r,0)$ be the coordinates of the RF AP, and $(x_u, y_u, z_u)$ be the coordinates of the user. Here, we set $z_u = -d_v$, while the user is uniformly located within the $x-y$ plane representing the VLC coverage area. Here, we set the vertical distance between the transmitting and receiving planes to $d_v = 2.5$~m.
\begin{table}[t]
\caption{Simulation Parameters}
\begin{center}
\begin{tabular}{l|c}\hline\hline
\multicolumn{2}{c}{{\bf RF System}}\\
\hline
Channel bandwidth, $B_{r}$ & $20$~MHz\\ 
Average emitted power, $P_{r} $ &  $10$~mW\\
Path loss exponent,  $\varrho$  & $1.6$ \\
Rician factor, $K$ & $5$~dB\\ 
Log-normal standard deviation, $\sigma$ &  $1.8$~dB\\ 
Noise power spectral density, $N_r$  &  $-114$~dBm/MHz\\
\hline
 \multicolumn{2}{c}{{\bf VLC System}}\\
\hline
PD field of view (FOV), $\psi_C$  & $ 90^{\circ}$ \\
Average emitted power, $P_v $ &  $9$~W\\
PD physical area, $A$ & $ 1 \,\text{cm}^2$\\ 
Modulation bandwidth, $B_v$ & $40$~MHz\\ 
PD responsivity, $\alpha$ &  $0.53$~A/W\\ 
Refractive index, $n$ & $1.5$\\ 
Optical filter gain, $D(\psi)$ & $1$\\ 
Noise power spectral density, $N_v$ & $10^{-21}$~$\text{A}^2/$Hz\\
Elect./opt. power conversion, $\varsigma$ & $3$\\
\hline\hline
\end{tabular}
\end{center}
\label{tab_1}
\end{table}%

\begin{figure}
\centering
\includegraphics[width=\figsizee\textwidth]{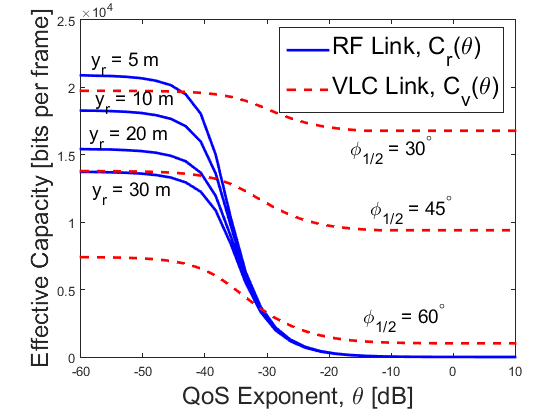}
\caption{Effective capacity of both RF and VLC links as a function of the QoS exponent, while considering different values of the distance $y_r$ and the LED viewing angle $\phi_{1/2}$.}
\label{EC_Hyp_theta_yr_phi}
\vspace{-0.5 cm}
\end{figure}

In \figurename~\ref{EC_Hyp_theta_yr_phi}, we illustrate the  effective capacity of both RF and VLC links with respect to the QoS exponent, $\theta$. We show the effective capacity of the RF link when the RF AP is located at different locations, expressed in terms of the distance $y_r \in \{5,10,20,30\}$ m, while we set $\phi_{1/2} = 45^{\circ}$. On the other hand, we plot the effective capacity of the VLC link considering different values of the LED viewing angle, 
\textit{i.e.}, $\phi_{1/2} \in \{30^{\circ},45^{\circ},60^{\circ}\}$. We immediately observe that increasing the QoS exponent reduces the supported arrival rate at the buffer when either link is used for the data transmission. However, increasing $\theta$ has a higher impact on the RF link, such that the link performance degrades rapidly and approaches zero after a certain value of $\theta$. On the other hand, the VLC link has a better immunity against the impact of increasing $\theta$, especially at smaller LED angle views (\textit{e.g.}, $\phi_{1/2} = 30^{\circ}$). These observations can be explained to the higher randomness nature of the RF link, which limits its ability to support stricter constraints. Nevertheless, such a stochastic nature can be beneficial at looser QoS constrains, as clearly seen in \figurename~\ref{EC_Hyp_theta_yr_phi} for $\theta <-35$ dB. We further notice that decreasing the LED viewing angle can significantly enhance the VLC link performance, which is expected since the transmitted power is being focused in a more confined area. Finally, we remark that the considered LED viewing angles in this figure, \textit{i.e.}, $\phi_{1/2} \in \{30^{\circ},45^{\circ},60^{\circ}\}$, guarantee having illuminance span of $\frac{E_{\text{max}}}{E_{\text{min}}} \geq \{3,5.6,16\}$, respectively. Notice that we generally have $\frac{E_{\text{max}}}{E_{\text{min}}} \geq 1$, such that the values closer to one correspond to stricter illumination needs over the entire cell.

\begin{figure}
\centering
\includegraphics[width=\figsizee\textwidth]{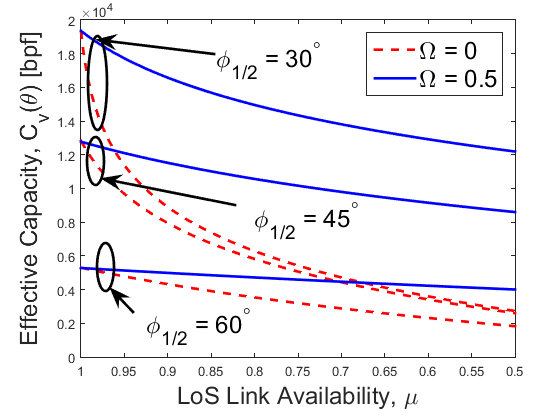}
\caption{Effective capacity of the VLC link as a function of the LoS availability for different values of the LED viewing angle $\phi_{1/2}$ and the rate ratio $\Omega$. Here, $\theta = -30$ dB. bpf = \textit{bits per frame}.}
\label{EC_VLC_mu}
\vspace{-0.5 cm}
\end{figure} 

The effect of the LoS disturbance (blockage) on the VLC link performance is depicted in \figurename~\ref{EC_VLC_mu}, where we plot the effective capacity of the VLC link as a function of the LoS probability $\mu$, and considering different values of the rate ratio $\Omega = \{0,0.5\}$. Notice that $\Omega = 0$ means that data transmission is not possible over the VLC link when the LoS access is blocked. We clearly observe that the VLC performance is highly affected by the LoS blockage, even when data transmission is maintained through the diffuse (NLoS) components, \textit{i.e.}, for $\Omega >0$. However, the effect of the LoS blockage is reduced at wider cell areas, \textit{e.g.}, for $\phi_{1/2} = 60^{\circ}$. This can be explained since the channel randomness due to the random user location is getting more dominant at wider cell areas. 

\begin{figure}
\centering
\subfigure[TDMA]{\includegraphics[width=\figsizee\textwidth]{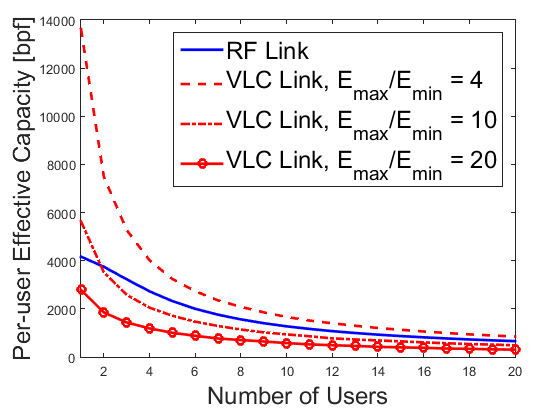}}\\
\subfigure[FDMA]{\includegraphics[width=\figsizee\textwidth]{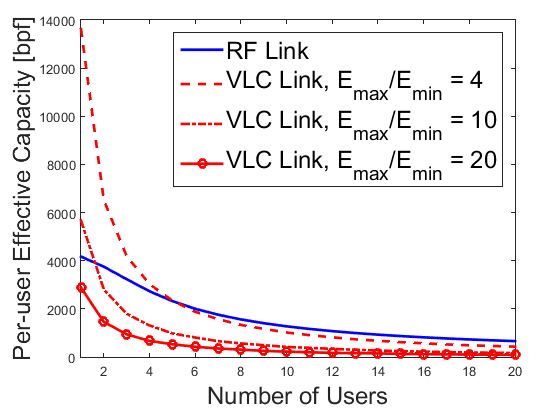}}
\caption{Effective capacity of both RF and VLC links as a function of the number of users (or equivalently the per-user allocated resources) and considering time-division and frequency division multiple access schemes, \textit{i.e.}, TDMA and FDMA, respectively. The effective capacity of the VLC link is shown for different illumination needs. Here, we set $y_r = 20$~m and $\theta = -30$~dB. bpf = \textit{bits per frame}.}
\label{EC_Hyp_num_users}
\vspace{-0.65 cm}
\end{figure}

In typical indoor scenarios, multiple users exist within the coverage area of the VLC AP and they share the same communication resources, \textit{i.e.}, power, time, and bandwidth. Therefore, different multiple access schemes are normally applied to share these resources among users. Herein, we consider the well-known time-division and frequency-division multiple access schemes, \textit{i.e.}, TDMA and FDMA, respectively, such that both the RF and VLC APs perform the same scheme. In TDMA, each user is allocated the total power and bandwidth, whereas the transmission period is \textit{equally} allocated among all users. In FDMA, each user can use the whole frame duration for transmission, while the available power and bandwidth\footnote{From practical perspectives, employing TDMA requires a good synchronization between the access point controller and each receiver, which can be done using a dedicated control channel. On the other hand, in FDMA each user should be designed to operate in a specific spectrum according to the allocated sub-channels. This can be realized by  integrating several sub-photodetectors in each receiver, such that each  sub-photodetector is designed to receive the data over a certain frequency (wavelength) range. Nevertheless, such practical aspects are beyond the scope of this paper.} are \textit{equally} divided among users. Assuming that each user is uniformly located within the VLC coverage area, we show the per-user effective capacity of RF and VLC links in \figurename~\ref{EC_Hyp_num_users} with respect to the number of users (or equivalently the per-user allocated resources). Also, the VLC link performance is shown for different illumination needs, expressed in terms of the span ratio $E_{\text{max}}/E_{\text{min}}$. Herein, we assume that all users have the same QoS needs and that $\theta = -30$~dB for all of them. 

For a low number of users, the VLC link generally outperforms the RF link, which agrees with the results obtained in \figurename~\ref{EC_Hyp_theta_yr_phi} for $\theta = -30$~dB. On the contrary, the RF link provides a better per-user performance when the number of users increases with both multiple access schemes, which means that the VLC performance is more sensitive to the allocated resources. Moreover, we also observe that the RF link has the same performance with both TDMA and FDMA schemes when the relevant resources are equally allocated among users, whereas the VLC link has a worse performance when the FDMA scheme is employed. This is because the signal-to-noise ratio of the RF link is linearly proportional with respect to the average power, and hence allocating the available power and bandwidth equally among all users keep the same signal-to-noise ratio value. On the other hand, the signal-to-noise ratio of the VLC link is a quadratic equation with respect to the average power, as seen in \eqref{SNR_vlc}, thus it is more sensitive to changes in the allocated power and/or bandwidth. This is indeed an important aspect when designing such a hybrid network, which reveals that the dimming issues in LED-based APs should be carefully handled.

\begin{figure}
\centering
\subfigure[RF Channel, $\phi_{1/2} = 45^{\circ}$]{\includegraphics[width=\figsizee\textwidth]{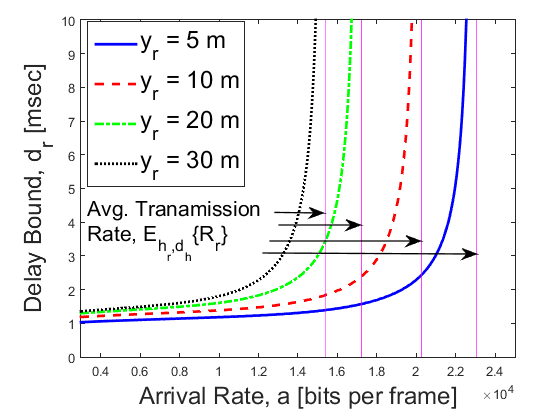}}\\
\subfigure[VLC Channel]{\includegraphics[width=\figsizee\textwidth]{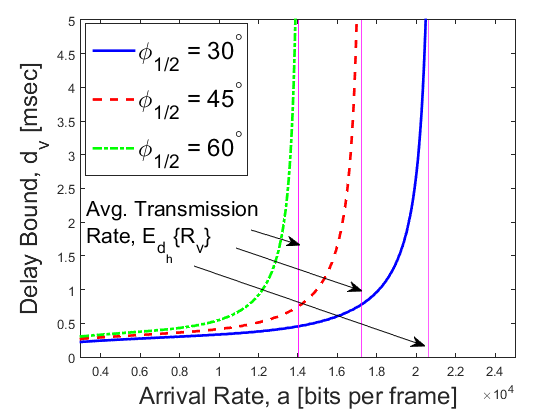}}
\caption{Delay Bounds of RF and VLC channels as a function of the data arrival rate and for different values of $y_r$ and $\phi_{1/2}$, respectively. Here, $\varepsilon = 10^{-6}$.}
\label{fig:Delay}
\end{figure}

Finally, in \figurename~\ref{fig:Delay} we show the delay bounds of RF and VLC channels with respect to the data arrival rate $a$. We observe that the RF link can perform better that the VLC in some settings, \textit{e.g.}, when the RF AP is located at $y_r = 5$~m, in terms of the supported arrival rates, \textit{i.e.}, the curve is shifted to the right. Nevertheless, the VLC link has lower delay bounds, \textit{i.e.}, the curves are shifted downwards, in all considered settings. Further, we notice that the delay bounds of both links increase asymptotically as the arrival rate approaches the average transmission rate of the link, which is expected since the systems turns to be unstable and data packets are expected to be buffered for longer periods.

\section{Conclusions}
\label{sec:conc}
In this paper, we have investigated the effective capacity of a hybrid RF/VLC system, which is subject to QoS constrains in the form of limits on the buffer overflow and delay violation probabilities. We have initially formulated the effective capacity of both RF and VLC links with respect to the channel conditions and user distribution. Then, we have provided a closed-form approximation for the effective capacity of the VLC link under the assumption of uniform user distribution within the cell. Non-asymptotic bounds on the buffering delay corresponding to both RF and VLC channels have been also provided. Simulation results have showed that the RF link is more beneficial at looser constraints and/or when more users are located in the VLC cell. On the other hand, the VLC link can support stricter constraints and/or less number of users. In addition, we have showed that while RF links can perform better in terms of the supported arrival rates in some settings, VLC links have a better delay performance, \textit{i.e.}, lower delay bounds. 
We have further displayed the impacts of illumination needs and the LoS blockage on the VLC performance.

\bibliographystyle{IEEEtran}
\bibliography{references}

\end{document}